# Novel g-computation algorithms for time-varying actions with recurrent and semi-competing events

Alena S. D'Alessio[1,2], Lucas M. Neuroth[1], Jessie K Edwards[1], Chantel L. Martin[1,2], Paul N Zivich[1]

[1]Department of Epidemiology, Gillings School of Global Public Health, University of North Carolina, Chapel Hill, NC

[2]Carolina Population Center, University of North Carolina, Chapel Hill, NC

## Abstract

**Background**: A core aspect of epidemiology is determining the impacts of potential public health interventions over time. With long follow-up periods, epidemiologists may need to consider semi-competing events, in which a terminal event, like death, precludes a non-terminal event, like hypertension. Time-varying confounding poses an additional challenge when studying time-varying interventions or actions. Existing methods do not simultaneously address semi-competing events and time-varying confounding.

**Methods**: We propose two novel g-computation algorithms for causal effects with semi-competing events and time-varying actions. To explore performance of our novel g-computation estimators, we conducted a Monte Carlo simulation study. We then applied our estimator to investigate how cigarette smoking prevention throughout young and middle adulthood might impact prevalent hypertension using data from Waves III (aged 18-26 years) - VI (aged 39-51 years) of the National Longitudinal Study of Adolescent to Adult Health.

**Results**: Our simulations show that the novel g-computation estimators had little bias and appropriate confidence interval coverage. They outperformed existing alternative estimators across sample sizes. In the illustrative application, the novel estimator identified a small reduction in prevalence of hypertension and risk of death in midlife had all cigarette smoking been prevented across follow-up compared to the observed smoking patterns.

**Conclusion:** As long-running cohorts progress in age, death within the study sample will become an increasing concern for studies of aging-related outcomes, life course analyses, and investigations into chronic disease development. Our novel g-computation estimators provide a simultaneous solution.

A core aspect of epidemiology is determining the impacts of different courses of actions in public health. In settings with long-term follow-up, semi-competing events, terminal events (e.g., death) that preclude the event of interest (e.g., hypertension; referred to as an intermediate, or transient, event in relation to the terminal event) are common.[1–4] Semi-competing events are distinct from competing events, where each event precludes the other (e.g., cause-specific death). When analyzing data with semi-competing events, simply censoring those who experience a terminal event complicates the interpretation and can make the analysis less useful for public health decision making.[5] For example, censoring implicitly imposes an intervention that prevents deaths without altering the hazard of hypertension. Accounting for semi-competing events is further complicated when actions under consideration are time-varying. Here, subsequent time-varying confounding, where a variable is affected by prior exposure while also affecting subsequent exposure and outcome, can also produce bias.[6] Both issues present substantial challenges to life course and aging research, where follow-up often spans decades and exposures vary over time.

Janvin et al. review different estimands, identification assumptions, and estimators for semi-competing event settings,[7] including the elimination of competing events and separable effects.[7–9] Separable effects focus on the decomposition of the effect of an action into distinct pathways, one for the recurrent event and one for the competing events.[10,11] Depending on the action of interest, isolation of the effects from the recurrent and terminal event may not be plausible. Rather than trying to decompose the mechanism, one might be interested in the so-called "total effect," which is the effect on both the recurrent and competing event. How an action affects both event types may be particularly relevant for public health decision making, as this would reflect how changing currently feasible actions would modify the outcomes in a population. Finally, prior work has primarily focused on settings where the action is only modified at baseline.[7,11] An exception is Díaz et al.,[12] which considered the total effect of policies that shift probabilities of time-varying actions with competing events. However, this work did not consider recurrent events.

Our work leverages a multistate illness-death-recovery model to account for changes in disease state over time.[2,13,14] Figure 1 presents a multistate model for death and hypertension where individuals are allowed to freely transition between hypertension status across follow-up, but death is an absorbing state.[15–18] We combine this multistate outcome model and the parametric g-formula to estimate the total effect of time-varying actions on all outcome types.[19,20] G-computation, an estimator of the g-formula and an established method to account for time-varying confounding,[19–23] has been extensively used to address competing events with baseline or time-varying actions.[20,24–27] This method can be implemented by either fitting a set of parametric regression models for all time-varying variables (standard g-computation);[20,22] or a series of nested outcome models (iterated conditional expectation (ICE) g-computation).[23] G-computation with reccurent and semi-competing events has been primarily limited to baseline actions.[14] Therefore, extensions of g-computation for estimation of the total effect in semi-competing event settings for time-vary actions are needed.

In this paper, we propose novel standard and ICE g-computation algorithms to estimate causal effects of time-varying actions in the semi-competing event setting. We describe algorithms for both estimators and provide results from a simulation study showing estimator performance. Using data from the National Longitudinal Study of Adolescent to Adult Health (Add Health), we illustrate the proposed approach in the context of cigarette smoking prevention and prevalent hypertension. To ease adoption, we provide code in R and Python for the proposed novel g-computation estimators (github.com/asdalessio/publications-code).

## Methods

Let $A_{i,k}$ denote a binary action (e.g., exposure, treatment, intervention) and $L_{i,k}$, denote a set of covariates measured at the discrete time $k \in \{0, \dots, \tau\}$ for individual $i$. Further, let $C_{i,k}$ denote whether an individual was censored, either administratively or due to loss-to-follow-up. Hereafter, we assume that once individuals are lost to follow-up, they remain censored for all following time points. Let $Y_{i,k}$ denote a multistate outcome that captures both a recurrent, intermediate event (e.g., hypertension) and terminal event (e.g., death). Let $Y_{i,k} \in \{1,2,3\}$ where $Y_{i,k} = 1$ indicates state with no observed intermediate or terminal event at time $k$ (e.g., individual is alive and not hypertensive), $Y_{i,k} = 2$ indicates state with observed intermediate event (e.g., individual is hypertensive), and $Y_{i,k} = 3$ indicates state with observed terminal event (e.g., individual is dead). Here, $Y_{i,k}$ is observed regardless of $Y_{i,k-1}$ when $Y_{i,k-1} \neq 3$ (e.g., the individual was alive at the previous time). If $Y_{i,k-1} = 3$, $L_{i,k}, A_{i,k}, Y_{i,k}$ is set to missing at all following times. At baseline, all individuals are assumed to be alive (i.e., $Y_{i,0} = \{1,2\}$). Data are assumed to be measured in the following temporal order: $L_0 \rightarrow A_0 \rightarrow C_1 \rightarrow Y_1 \rightarrow L_1 \rightarrow A_1 \rightarrow \cdots \rightarrow C_\tau \rightarrow Y_\tau$. Finally, overbars are used to denote variable histories (e.g., $\bar{A}_{i,k} = (A_{i,0}, A_{i,1}, \dots, A_{i,k})$) and we allow $\bar{L}_{i,k-1}$ to include $\bar{Y}_{i,k-1}$.

We define an action plan as a deterministic function that sets the sequence of actions to fixed values across follow-up (i.e., $\bar{a}^*_{\tau-1} = a^*_0, a^*_1, \dots, a^*_{\tau-1}$). Let $Y^{\bar{a}_{\tau-1}}_{i,\tau}$ denote the potential outcome under plan $\bar{a}_{\tau-1}$ for individual $i$ . In settings with a competing event, causal parameters for a single outcome state (i.e., $I\left(Y^{\bar{a}_{\tau-1}}_{i,\tau} = 2\right)$) are no longer sufficient to inform decisions. For example, one might trivially prevent hypertension by inducing death. To avoid these concerns, the estimand in competing event settings is instead a collection of contrasts for the outcomes between two action plans $\bar{a}_{\tau-1}$ and $\bar{a}'_{\tau-1}$. Because there are three states, our estimand is the following pair, or vector, of proportion differences

$$\psi(\tau) = \begin{bmatrix} \psi_2(\tau) \\ \psi_3(\tau) \end{bmatrix} = \begin{bmatrix} E\left[I\left(Z^{\bar{a}_{\tau-1}}_{\tau} = 2\right)\right] - E\left[I\left(Z^{\bar{a}'_{\tau-1}}_{\tau} = 2\right)\right] \\ E\left[I\left(Z^{\bar{a}_{\tau-1}}_{\tau} = 3\right)\right] - E\left[I\left(Z^{\bar{a}'_{\tau-1}}_{\tau} = 3\right)\right] \end{bmatrix}$$

where $E[\cdot]$ is the expected value function and $Z_{i,k}$ represent the state individual $i$ is in at time interval $k$ such that $Z_{i,k} = 3$ if individual $i$ experienced the terminal event at any time $j < k$, otherwise, $Z_{i,k} = Y_{i,k}$. Therefore, we are interested in the average causal effects of $\bar{A}_k$ on the recurrent, intermediate state (prevalence difference) and terminal state (risk difference). The prevalence difference for the referent, no event state ($\psi_1(\tau)$) can be inferred from $\psi_2(\tau)$ and $\psi_3(\tau)$ given $\sum_{q=1}^{3} \psi_q(\tau) = 0$.

### *Identification*

We now describe a sufficient set of assumptions under which $\psi(\tau)$ is identified (i.e., expressable in terms of the observed data). These assumptions are presented formally in Table 1. First, causal consistency links the potential variables under the observed actions to the observed variables. Here, causal consistency implies no interference between units and action-variation irrelevance.[28] Next, we assume time-varying action exchangeability (i.e., conditional independence) between the potential outcomes and actions conditional on prior observed covariates and actions. Additionally, we allow for informative loss-to-follow-up through an additional time-varying censoring exchangeability assumption conditional on prior covariates and actions. Each exchangeability assumption comes with a corresponding positivity assumption. Comprehensive discussion of these assumptions in related settings has been given elsewhere.[14,20,29,30] Under these assumptions, it follows that

$$\psi(\tau) = \begin{bmatrix} \mu_{2,\tau}(\bar{a}_{\tau-1}) - \mu_{2,\tau}(\bar{a}'_{\tau-1}) \\ \mu_{3,\tau}(\bar{a}_{\tau-1}) - \mu_{3,\tau}(\bar{a}'_{\tau-1}) \end{bmatrix}$$

where $\mu_{z,\tau}(\bar{a}_{\tau-1}) = E[E\{\dots E[I(Z_\tau = z) \mid \bar{A}_{\tau-1} = \bar{a}_{\tau-1}, \bar{L}_{\tau-1}, Y_{\tau-1} \neq 3, C_{\tau-1} = 0] \dots \mid A_0 = a_0, L_0\}]$. Proof of this equality is presented in eAppendix A.

*Estimation*

For estimation in the semi-competing events setting with time-varying confounding, we adapt previously described g-computation algorithms for time-varying confounding with repeated measures and for competing events.[20,23] For our proposed standard and ICE g-computation algorithms, we assume that the data is in the "wide" structure with one row per individual $i$ and one column per time-unit $k$ for the action ($A_{i,k}$), confounders ($L_{i,k}$), and outcome ($Y_{i,k}$).

*Algorithm - Standard g-computation*

1. For each follow-up $k \in \{1, \dots, \tau\}$, fit regression models for $Y_{i,k}$ and each variable in $L_{i,k}$ conditional on $\bar{A}_{i,k-1}$, $\bar{L}_{i,k-1}$. A multinomial logistic model for $Y_{i,k}$ for all individuals free of the terminal event at $k$ (i.e., $Y_{i,k} \neq 3$) is fit. For $L_{i,k}$, the chosen model is based on the variable (e.g., logistic model for a binary $L_{i,k}$).
2. Sample with replacement $L_0$ from the observed data $B$ times, where $B$ is a large number (e.g., 100,000).
3. For each follow-up $k \in \{1, \dots, \tau\}$,
    a. If $k = 1$ or the predicted value of $Y_{i,k-1}$, denoted $\check{Y}_{i,k-1}$, is not the terminal state (i.e., $\check{Y}_{i,k} \neq 3$), then
        i. Set the action plan of interest (i.e., $\bar{a}^*_{k-1} = 1$).
        ii. Generate predicted probabilities of $\widehat{\boldsymbol{Y}}_{i,k}$ under the action plan, $\bar{a}^*_{k-1}$, and $\bar{L}_{i,k-1}$, such that the predicted probabilities are structured as an $1 \times 3$ matrix of probabilities for each possible outcome state.
        iii. Resolve the predicted probabilities by drawing from a multinomial distribution to generate $\check{Y}_{i,k}$.
        iv. If $\check{Y}_{i,k} = \{1,2\}$, generate predicted values for $L_{i,k}$ under $\bar{a}^*_{k-1}$ and $\bar{L}_{i,k-1}$. Else, set $L_{i,k}, a^*_k$ and $Y_{i,k+1}$ to missing.
    b. Else, set $L_{i,k}, a^*_k$ and $Y_{i,k+1}$ to missing.
4. Using the simulated data from Step 3, create a composite outcome $\check{Z}_{i,\tau}$.
5. Take the proportion of each level of $\check{Z}_{i,\tau}$ such that $\frac{1}{B}\sum_{i=B}^{B} I(\check{Z}_{i,\tau} = 2)$ is the proportion of intermediate events and $\frac{1}{B}\sum_{i=B}^{B} I(\check{Z}_{i,\tau} = 3)$ is the proportion of terminal events.

To estimate $\psi(\tau)$, steps 2-5 are repeated with an alternate action plan, $\bar{a}'_{k-1}$, and then the differences in proportions are contrasted, yielding a vector of the prevalence difference for the intermediate state and risk difference for the terminal state.

*Algorithm - ICE g-computation*

1. Fit a multinomial logistic regression model for $Y_{i,\tau}$ conditional on $\bar{A}_{i,\tau-1}$, $\bar{L}_{i,\tau-1}$ for all individuals uncensored at $\tau$ and free of the terminal event at $\tau - 1$ (i.e., $C_{i,\tau} = 0$ and $Y_{i,\tau-1} \neq 3$).
2. From the estimated regression model, generate a vector of predicted probabilities for each outcome state under the action plan, $\bar{a}^*_{\tau-1}$, and $\bar{L}_{i,\tau-1}$ for all individuals uncensored and free of the terminal event at $\tau - 1$ (i.e., $C_{i,\tau-1} = 0$ and $Y_{i,\tau-1} \neq 3$) denoted $\widehat{\boldsymbol{Y}}_{i,\tau-1}$.
3. For $k \in \{\tau - 1, \tau - 2, \dots, 1\}$
    a. If $Y_{i,k} = 3$ (the terminal event), replace $\widehat{\boldsymbol{Y}}_{i,k}$ with $[0,0,1]$.
    b. Fit a multinomial logistic regression model for $\widehat{\boldsymbol{Y}}_{i,k}$ conditional on $\bar{A}_{i,k-1}, \bar{L}_{i,k-1}$ for all individuals uncensored and free of the terminal event at $k$ (i.e., $C_{i,k} = 0$ and $Y_{i,k} \neq 3$).
    c. Generate $\widehat{\boldsymbol{Y}}_{i,k-1}$ under $\bar{a}^*_{k-1}$ and $\bar{L}_{i,k-1}$ for all individuals uncensored and free of the terminal event at $k - 1$.
4. For each outcome state $\widehat{\boldsymbol{Y}}_{i,0}$, take the mean across all $n$ individuals as the estimate of proportion in each state under $\bar{a}^*_{\tau-1}$.

Again, $\psi(\tau)$ is estimated by repeating the algorithm under the alternate action plan, $\bar{a}'_{\tau-1}$, and contrasting the proportions under the two plans.

For either g-computation algorithm, the variance and point-wise Wald-type confidence intervals can be estimated using a non-parametric bootstrap procedure.[31,32] Alternatively, the variance for ICE g-computation can be estimated using the empirical sandwich variance estimator to avoid the computational burden of the bootstrap.[23]

## Simulation

To explore some properties of our proposed estimators, we conducted a Monte Carlo simulation study. We compared the proposed estimators to two alternatives. The first alternative estimator is a multistate g-computation estimator that only considers the action at baseline (i.e., an intent to treat estimator for a per protocol parameter). The second alternative estimator is an ICE g-computation estimator that accounts for time-varying confounding but censors the terminal event. These alternative estimators capture common conceptual approaches: addressing semi-competing events but not time-varying actions, or accounting for time-varying actions but ignoring terminal events. The parameters of interest for the simulations was $\psi(3)$, the proportion differences for the intermediate event ($Z^{\bar{a}_{\tau-1}}_{i,\tau} = 2$) and the terminal event ($Z^{\bar{a}_{\tau-1}}_{i,\tau} = 3$) had everyone had the action ($\bar{a}_{i,3} = (1, 1, 1)$) compared to no one had the action ($\bar{a}_{i,3} = (0, 0, 0)$). Simulations were conducted in R Statistical Software (version 4.4.0).

*Data generating mechanism*

We briefly describe the data generating mechanism, with details provided in eAppendix B. We generated data for $L$, $A$, and $Y$ at three follow-up times. Here, $L_{i,k}$ is a dichotomous, time-varying confounder that depends on $L_{i,k-1}$ and $A_{i,k-1}$. Similarly, $A_{i,k}$ depended on $L_{k-1}$ and $A_{k-1}$. Potential outcomes follows the multistate structure previously described (i.e., 2 denotes the intermediate event, 3 denotes the terminal event), and were generated from a multinomial distribution: $\frac{\exp(\eta_{i,j})}{\sum_{k=0}^{2}\exp(\eta_{i,k})}$ where $j \in \{1,2,3\}$ and the linear function $\eta_{i,0} = 1$ is the reference while $\eta_{i,2}$ and $\eta_{i,3}$ are defined for each time point, $k$.

We generate the observed values of $L$ and $Y$ from their simulated potential outcomes via causal consistency. Censoring at time $k$, $C_{i,k}$, was induced among those alive at the start of the interval and depended on only $A_{k-1}$. When censored, $L_{i,k}, A_{i,k}, Y_{i,k}$ are unobserved for all subsequent times.

*Simulation parameters and evaluation metrics*

We considered two sample sizes $n \in \{500, 2000\}$ with 2,000 iterations each. To estimate standard errors, we used the non-parametric bootstrap with 500 repetitions. Each estimator was evaluated using the following metrics: bias, empirical standard error (ESE), RMSE (root mean standard error), average standard error (ASE), standard error ratio, and 95% confidence interval (CI) coverage.[33] For estimators that did not estimate one of the proportion differences in $\psi(\tau)$, the metrics for the ineligible proportion difference were undefined.

*Simulation Results*

Simulation results are presented in Table 2. Both proposed g-computation estimators were approximately unbiased for both sample sizes. Regarding precision, the proposed estimators were similar in terms of ESE. The bootstrap was unbiased for variance estimation, with a SER near 1 and 95% CI coverage near 0.95 for both proposed estimators.

Both alternative estimators considered were biased irrespective of sample size. Further, the bias for each estimator was in opposing directions. While the SER was near 1 for the alternative estimators, bias led to 95% CI coverage below 0.95. The smaller bias for alternative estimator 2 relative to the other alternative estimator is likely due to the relatively low proportion of the terminal event in the population (~13% regardless of the action plan). For each individual parameter in $\psi(\tau)$, the g-computation estimators had the lowest RMSE.

## Application

*Methods*

To illustrate the implementation of the proposed estimators, we estimate the effect of preventing cigarette smoking throughout young and middle adulthood on later-life hypertension using Add Health. Add Health is a longitudinal cohort study of adolescents beginning in grades 7-12 during the 1994-1995 academic school year.[34] Approximately 90,000 students from a nationally representative stratified random sample of middle and high schools took an in-school survey (Wave I). A random subsample of students ($n = 20{,}745$) were selected to participate in longitudinal in-home interviews with follow-up in 1996 (Wave II, grades 8-12, $n = 14{,}738$), 2001-2002 (Wave III, ages 18-26, $n = 15{,}107$), 2008-2009 (Wave IV, ages 24-32, $n = 15{,}701$), 2016-2018 (Wave V, ages 33-43, $n = 12{,}300$), and 2022-2025 (Wave VI, ages 39-51, $n = 11{,}979$). Cardiovascular measures (e.g., systolic and diastolic blood pressure) were introduced in Wave IV. Wave V restricted biomarker measurement to a subsample of participants ($n = 5{,}361$).

Our illustrative example uses data from Waves III-VI ($n = 14{,}824$ at Wave III). Individuals with Wave IV data who were not included in the Wave V sub-sample were administratively censored at Wave V and VI. As in Zivich et al. 2024,[23] individuals with Wave III follow-up were excluded from the analysis if they were younger than age 13 or older than age 18 at Wave I for data sparsity and to make participants more homogeneous in age ($n = 830$). Participants were also excluded if they reported ever taking a medication for a heart condition at Wave III (indicating a potential underlying health condition) ($n = 62$). The final analytic sample size was 13,909 participants.

Hypertension and death were assessed at Waves IV-VI. Participants with systolic blood pressure ≥140 mmHg or diastolic blood pressure ≥90 mmHg were considered hypertensive for the given follow-up. Death was captured using the final disposition of respondents for each study follow-up. Hypertension and death were coded as a single categorical outcome to align with the previously defined multistate outcome variable: $Y_{i,k} \in \{1,2,3\}$ where $Y_{i,k} = 1$ indicates no observed hypertension or death at time $k \in \{1, 2, 3\}$, $Y_{i,k} = 2$ indicates observed hypertension, and $Y_{i,k} = 3$ indicates observed death. The estimand was the proportion differences for hypertension ($\psi_2(t)$) and death ($\psi_3(t)$) at Wave VI ($t = 3$) had none of the sample ever been current smokers at Waves III, IV, and V ($\bar{a}_{i,2} = (0, 0,0\ )$) compared to the naturally observed smoking patterns over follow-up. With notation, the estimand was $\psi_2(3) = E\left[I\left(Z_3^{\bar{a}_2} = 2\right)\right] - E[I(Z_3 = 2)]$ and $\psi_3(3) = E\left[I\left(Z_3^{\bar{a}_2} = 3\right)\right] - E[I(Z_3 = 3)]$.

Time-varying confounders included height, weight, exercise frequency, alcohol consumption, self-rated health, prior diagnosis of hypertension, prior smoking history, educational-attainment, and health insurance coverage. Baseline confounders included gender, race, ethnicity, and age at Wave I as well as ever trying a cigarette at Wave III. Variable definitions are provided in the eAppendix C, eTable1. If a participant had missing covariates in a given follow-up wave, all data were set to missing in that wave and future waves. This ensured monotonic missingness across time and avoided imputation for simplicity of the illustrative example.

For estimation of $\psi$, the proposed ICE g-computation estimator was compared against the two naïve estimators. We did not implement the proposed standard g-computation algorithm due to the computational burden of modeling the time-varying confounders. Height and weight were modeled with restricted quadratic splines, and all other confounders were modeled as disjoint indicators. Additional information on model specifications is provided in the supplemental material. Variance was estimated using a non-parametric bootstrap with 500 repetitions.

*Results*

Descriptive statistics for the analytic sample are presented by follow-up wave in eAppendix C, eTable 2. Wave III included 13,909 participants, 32% of whom reported being current cigarette smokers. Between Waves III-IV, $n = 2{,}090$ (18%) participants had developed hypertension, $n = 77$ (1%) died, $n = 2{,}346$ (17%) were censored. At Wave V, a smaller sub-sample of participants underwent hypertension measurement, leading to $n = 7{,}535$ (54%) participants in the analytic sample being censored. Between Waves IV-V, $n = 669$ (17%) of participants had developed hypertension, $n = 130$ (3%) died. Between Waves V-VI, $n = 459$ (18%) participants had developed hypertension, $n = 21$ (1%) died, $n = 549$ (18%) were censored. The small number of deaths between Waves V-VI is likely due to shorter follow-up length between Waves V-VI (4 years) than Waves IV-V (7 years).

Results for the applied example across Waves are presented in Figure 2. We estimated that the prevalence of hypertension would have been 1.1 percentage points (95% CI: -2.2, -0.1) lower on the absolute scale had all smoking been prevented (18.4% (95% CI: 16.6, 20.1) under intervention versus 19.5% (95% CI: 17.8, 21.1) under naturally occurring smoking pattern). Additionally, we found that the risk of death would have been 1.6 percentage points (95% CI: -2.3, -1.0) lower under the intervention (3.9% (95% CI: 3.1, 4.7)) compared to the natural pattern (5.5 95% CI: 4.7, 6.3). Both alternative estimators had less precision than our novel estimator as seen through wider confidence intervals in eTable 3. Alternative estimator 1 found a reduced prevalence difference of hypertension but an increased risk difference of death compared to our proposed estimator. Alternative estimator 2 found a similar prevalence difference of hypertension as our proposed estimator but assumes the risk of death is zero.

## Discussion

In this paper, we sought to combine aspects from causal inference and multi-state models to study the effects of time-varying actions on both recurrent and terminal events. Our proposed extension of g-computation appropriately accounts for time-varying confounders as well as estimates a multistate outcome with both intermediate and terminal disease states. Thus, the proposed estimators allow for the intermediate event prevalence to increase or decrease as individuals transition between non-terminal states during follow-up. This differs from common analytic approaches that assume individuals uniformly stay in a disease state after entrance, as is assumed in many analyses of chronic and aging-related disease outcomes with competing events. Our modeling of outcome states better aligns with dynamic changes in disease status. Our simulations show good performance of both proposed standard and ICE g-computation estimators. The novel estimators outperformed existing alternative estimators across different sample sizes. The applied example using Add Health data demonstrated how our estimator can be implemented. The proposed estimator allowed for identification of the intervention effect on both hypertension and death while also accounting for time-varying confounding, which neither alternative estimator could support.

Analyses leveraging longitudinal cohort studies are an important application of our proposed g-computation estimators. As long-running longitudinal research cohorts progress in age, death within the study sample will become an increasing concern for studies of aging-related outcomes, life course analyses, and investigations into chronic disease development. For example, the Add Health cohort entered Wave VI follow-up with 596 total deaths (3% of Wave I sample), and the proportion will increase at a growing rate as the cohort ages further into late adulthood. As we have shown in this paper, failure to account for the semi-competing risk of death induces non-ignorable bias. Analysts may consider imputing intermediate outcomes among those with the terminal event, but this is conceptually equivalent to censoring terminal events. Our proposed estimator is also relevant for other settings, like intercurrent events (e.g., discontinuation of treatment, death, etc.) in trials.[35] Here, the estimator could be used to estimate per-protocol effects with informative non-adherence and competing events.

It is important to note that our applied example is not intended to serve as an investigation into a public health intervention, but rather an illustration of our estimator. As such, there are several important limitations. First, confirmed deaths from National Death Index data were unavailable for Wave VI follow-up, therefore, we utilized disposition at study follow-up to identify deaths. This could have led to an undercounting of deaths. Second, although we accounted for censoring, we required participants to have complete data within each study follow-up to avoid complicating the illustration. Bias associated with complete-case analyses are well described and are not recommended in most epidemiologic investigations.[36] Third, we ignored other forms of tobacco consumption (e.g., vape, chew, snuff). Fourth, we did not consider anti-hypertensive medications usage or self-reported history of hypertension. Lastly, we did not apply the built-in Add Health sampling weights, but the proposed estimators could be adapted to weighted variations.

There are several ways this work can be extended. First, we focused on a scenario with a single intermediate event and a single terminal event, but additional states could be considered. For example, the intermediate state of hypertension could be separated into hypertensive stages, or the terminal state of death could be separated into cause-specific death. As the outcome states expand, the multi-state diagram presented in Figure 1 would take on a more complex structure; however, the proposed g-computation algorithms should be relatively straightforward to extend to these settings. Second, our example scenario only considered simplified deterministic plans for a binary action. Again, the proposed g-computation estimators could be extended to more complex deterministic plans, categorical or continuous actions, or stochastic (i.e., probabilistically assigned) plans.[37] Third, it remains of interest to develop inverse probability weighting estimators for semi-competing events with time-varying confounding. These estimators might be able to be further combined with ICE g-computation into a multiply-robust augmented inverse probability weighting estimator. Fourth, our presentation of confidence intervals in Figure 2 may understate the uncertainty, but this can be corrected for with confidence bands.[38] Finally, summary measures can be calculated, such as the proportion of time in each state and the restricted mean survival time.

Previous investigation into complex longitudinal data using g-computation was unable to address scenarios with simultaneously occurring time-varying actions and semi-competing events. Our proposed g-computation extension fills this important methodological gap. Researchers should consider utilizing the proposed estimators to more appropriately study the effects of interventions implemented across time while correctly accounting for semi-competing events in their sample.

Table 1. Sufficient identification assumptions for semi-competing event outcome with time-varying actions.

| **Assumption** | **Expression** | **Condition** | **Interpretation** |
|---|---|---|---|
| Causal consistency | $Z_{i,k} = Z_{i,k}(\bar{a}^*_{k-1})$<br>$L_{i,k} = L_{i,k}(\bar{a}^*_{k-1})$ | if $\bar{a}^*_{k-1} = \bar{A}_{i,k-1}$ | An action at any time $k$ produces the same effect. Also called treatment variation irrelevance. |
| Time-varying action exchangeability | $Z_k(\bar{a}^*_{k-1}) \coprod \bar{A}_{k-1} \mid \bar{A}_{k-2} = \bar{a}^*_{k-2}, \bar{L}_{k-1}$ | for $\bar{a}^*_{k-1}$ | An action at time $k$ is independent of the potential outcomes under each action plan, conditional on the time-varying action and confounding history. |
| Time-varying action positivity | $f(\bar{a}^*_k \mid \bar{a}^*_{k-1}, \bar{l}_{k-1}) > 0$ | for $\bar{a}^*_{k-1}, \bar{l}_{k-1}$ where $f(\bar{a}^*_{k-1}, \bar{l}_{k-1}) > 0$ | If the action plan has been followed through $k-1$ and given the time-varying confounding history, there is a non-zero probability that the action plan will be followed at time $k$. |
| Time-varying censoring exchangeability | $Z_k \coprod C_k \mid \bar{A}_{k-1}, \bar{L}_{k-1}, C_{k-1} = 0$ | for $C_k = 0$ | Censoring at time $k$ is independent of the outcome at that time, conditional on the time-varying action and confounder history. Also known as conditional non-informative censoring. |
| Time-varying censoring positivity | $Pr(C_k = 0 \mid \bar{a}^*_{k-1}, \bar{l}_{k-1}, C_{k-1} = 0) > 0$ | for $\bar{a}^*_{k-1}, \bar{l}_{k-1}$ where $f(\bar{a}^*_{k-1}, \bar{l}_{k-1}, C_{k-1} = 0) > 0$ | There is a non-zero probability of no censoring given the time-varying action and confounding history through $k-1$ and no censoring at the prior time. |

Notes: The overbar indicates variable history. $Z_{i,k}(\bar{a}^*_{k-1})$ is the potential outcome at time $k$ under the action plan $a^*$. $L_{i,k}(\bar{a}^*_{k-1})$ reflects the potential covariates at time $k$ under the action plan $a^*$ (can include $\bar{\bar{Z}}_{k-1}$). The probability density function if represented with $f$. All assumptions and conditions hold for all time points in follow-up.

Table 2. Simulation results for $t = 3$ comparing standard and ICE g-computation and alternative estimators.

| **Estimator** | **Bias** | **ESE** | **RMSE** | **ASE** | **SER** | **95% CI coverage** |
|---|---|---|---|---|---|---|
| | | | $n = 500$ | | | |
| Alternative estimator 1[b] | | | | | | |
| Intermediate event | 0.09 | 0.05 | 0.10 | 0.05 | 0.98 | 0.52 |
| Terminal event | 0.03 | 0.03 | 0.05 | 0.03 | 0.99 | 0.82 |
| Alternative estimator 2[c] | | | | | | |
| Intermediate event | -0.02 | 0.05 | 0.06 | 0.05 | 0.98 | 0.92 |
| Terminal event | - | - | - | - | - | - |
| Standard g-computation | | | | | | |
| Intermediate event | 0.00 | 0.05 | 0.05 | 0.05 | 1.00 | 0.95 |
| Terminal event | 0.00 | 0.03 | 0.03 | 0.03 | 1.01 | 0.95 |
| ICE g-computation | | | | | | |
| Intermediate event | 0.00 | 0.05 | 0.05 | 0.05 | 0.98 | 0.95 |
| Terminal event | 0.00 | 0.03 | 0.03 | 0.03 | 1.00 | 0.94 |
| | | | $n = 2{,}000$ | | | |
| Alternative estimator 1 | | | | | | |
| Intermediate event | 0.09 | 0.02 | 0.10 | 0.02 | 0.97 | 0.02 |
| Terminal event | 0.03 | 0.02 | 0.03 | 0.02 | 1.01 | 0.56 |
| Alternative estimator 2 | | | | | | |
| Intermediate event | -0.02 | 0.03 | 0.04 | 0.03 | 0.97 | 0.83 |
| Terminal event | - | - | - | - | | - |
| Standard g-computation | | | | | | |
| Intermediate event | 0.00 | 0.02 | 0.02 | 0.03 | 1.02 | 0.95 |
| Terminal event | 0.00 | 0.02 | 0.02 | 0.02 | 1.04 | 0.95 |
| ICE g-computation | | | | | | |
| Intermediate event | 0.00 | 0.02 | 0.02 | 0.02 | 0.98 | 0.94 |
| Terminal event | 0.00 | 0.02 | 0.02 | 0.02 | 1.00 | 0.95 |

ESE: empirical standard error; RMSE: root mean standard error; ASE: average standard error; SER: standard error ratio; 95% CI Coverage: 95% confidence interval coverage.

[a] Prevalence difference for the intermediate event and risk difference for the terminal event.

[b] G-computation estimator that accounts for semi-competing risks but only estimates the effect of the action at baseline.

[c] ICE g-computation estimator that accounts for time-varying confounding but treats terminal event as censoring (no estimation of terminal event risk).

Bias is calculated as the estimated proportion difference minus the true value averaged over the number of simulation iterations. ESE is the standard deviation of the estimates. RMSE is the square-root of bias squared plus ESE squared. ASE is the estimated standard errors averaged over the number of simulation iterations. SER is the ASE divided by the ESE. 95% CI coverage is the proportion of the estimated 95% CIs that contain the true value.

Figure 1. Multi-state model for hypertension and death

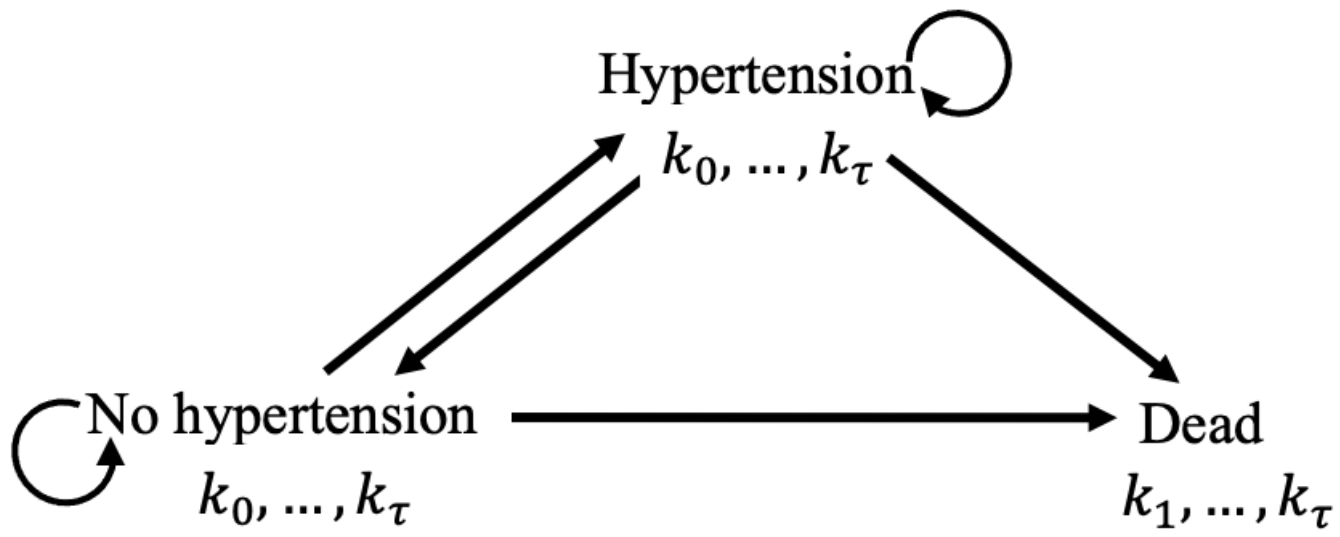

Arrows indicate possible flow of participants through three outcome states (no hypertension, hypertension, dead) over discrete time interval $k \in \{0, \dots, \tau\}$. At baseline, participants are in either the no hypertension or hypertension state. At the next follow-up, participants can remain in the same state or transition to one of the other two states. The death state is absorbing, such that once a participant dies, they remain in that state across subsequent follow-up.

Figure 2. Results at each Wave comparing effect of cigarette smoking prevention to the naturally observed smoking pattern from early adulthood to mid-adulthood on prevalent hypertension and incident death.

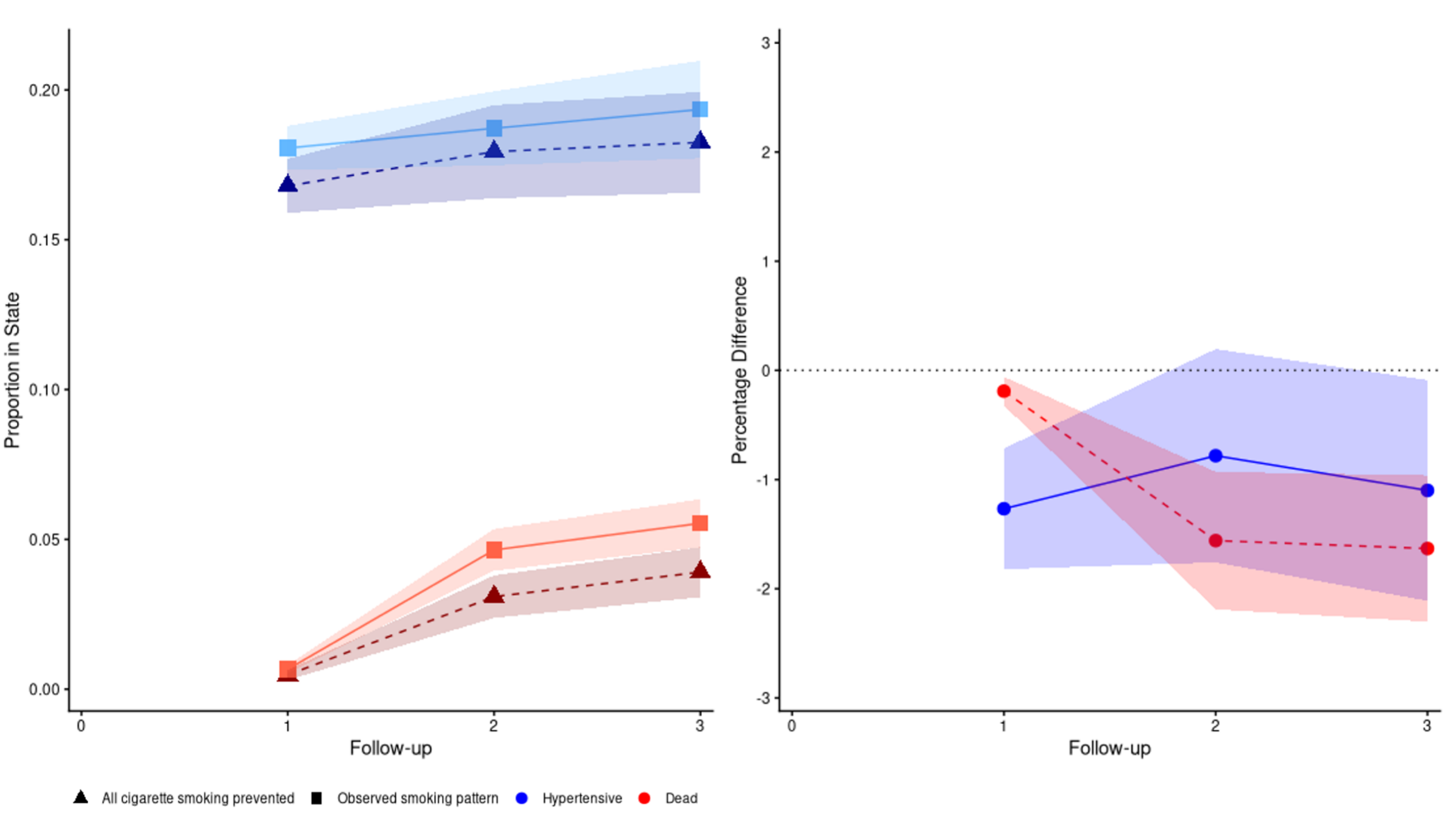

The left panel shows the proportion in each state under the two action plans. The right panel presents the prevalence difference for hypertension and risk difference for death contrasting the smoking prevention intervention and the naturally observed smoking pattern. Shaded regions denote the corresponding 95% confidence intervals.

## References

1. Fine JP, Jiang H, Chappell R. On Semi-Competing Risks Data. *Biometrika*. 2001;88(4):907-919.

2. Meira-Machado L, de Uña-Alvarez J, Cadarso-Suárez C, Andersen PK. Multi-state models for the analysis of time-to-event data. *Stat Methods Med Res*. 2009;18(2):195-222. doi:10.1177/0962280208092301

3. Wu R, Zhang Y, Bakoyannis G. Non-Parametric Estimation for Semi-Competing Risks Data With Event Misascertainment. *Stat Med*. 2025;44(3-4):e10332. doi:10.1002/sim.10332

4. Haneuse S, Lee KH. Semi-Competing Risks Data Analysis: Accounting for Death as a Competing Risk When the Outcome of Interest is Non-Terminal. *Circ Cardiovasc Qual Outcomes*. 2016;9(3):322-331. doi:10.1161/CIRCOUTCOMES.115.001841

5. Binder N, Schumacher M. Missing information caused by death leads to bias in relative risk estimates. *Journal of Clinical Epidemiology*. 2014;67(10):1111-1120. doi:10.1016/j.jclinepi.2014.05.010

6. Robins JM, Hernán MA, Brumback B. Marginal structural models and causal inference in epidemiology. *Epidemiology*. 2000;11(5):550-560. doi:10.1097/00001648-200009000-00011

7. Janvin M, Young JG, Ryalen PC, Stensrud MJ. Causal inference with recurrent and competing events. *Lifetime Data Anal*. 2024;30(1):59-118. doi:10.1007/s10985-023-09594-8

8. Young JG, Stensrud MJ, Tchetgen Tchetgen EJ, Hernán MA. A causal framework for classical statistical estimands in failure-time settings with competing events. *Stat Med*. 2020;39(8):1199-1236. doi:10.1002/sim.8471

9. Rojas-Saunero LP, Young JG, Didelez V, Ikram MA, Swanson SA. Considering Questions Before Methods in Dementia Research With Competing Events and Causal Goals. *American Journal of Epidemiology*. 2023;192(8):1415-1423. doi:10.1093/aje/kwad090

10. Stensrud MJ, Hernán MA, Tchetgen Tchetgen EJ, Robins JM, Didelez V, Young JG. A generalized theory of separable effects in competing event settings. *Lifetime Data Anal*. 2021;27(4):588-631. doi:10.1007/s10985-021-09530-8

11. Stensrud MJ, Young JG, Didelez V, Robins JM, Hernán MA. Separable Effects for Causal Inference in the Presence of Competing Events. *Journal of the American Statistical Association*. 2022;117(537):175-183. doi:10.1080/01621459.2020.1765783

12. Díaz I, Williams N, Hoffman KL, Hejazi NS. Author correction to: "causal survival analysis under competing risks using longitudinal modified treatment policies." *Lifetime Data Anal*. 2025;31(2):442-471. doi:10.1007/s10985-025-09651-4

13. Valeri L, Proust-Lima C, Fan W, Chen JT, Jacqmin-Gadda H. A multistate approach for the study of interventions on an intermediate time-to-event in health disparities research. *Stat Methods Med Res*. 2023;32(8):1445-1460. doi:10.1177/09622802231163331

14. Prosepe I, le Cessie S, Putter H, van Geloven N. Causal Multistate Models to Evaluate Treatment Delay. *Stat Med*. 2025;44(7):e70061. doi:10.1002/sim.70061

15. Dannenberg AL, Kannel WB. Remission of hypertension. The "natural" history of blood pressure treatment in the Framingham Study. *JAMA*. 1987;257(11):1477-1483. doi:10.1001/jama.257.11.1477

16. Schiavon CA, Bersch-Ferreira AC, Santucci EV, et al. Effects of Bariatric Surgery in Obese Patients With Hypertension. *Circulation*. 2018;137(11):1132-1142. doi:10.1161/CIRCULATIONAHA.117.032130

17. Xin X, He J, Frontini MG, Ogden LG, Motsamai OI, Whelton PK. Effects of alcohol reduction on blood pressure: a meta-analysis of randomized controlled trials. *Hypertension*. 2001;38(5):1112-1117. doi:10.1161/hy1101.093424

18. Fryar CD, Kit B, Carroll MD, Afful J. Hypertension Prevalence, Awareness, Treatment, and Control Among Adults Age 18 and Older: United States, August 2021-August 2023. *NCHS Data Brief*. 2024;(511).

19. Naimi AI, Cole SR, Kennedy EH. An introduction to g methods. *Int J Epidemiol*. 2017;46(2):756-762. doi:10.1093/ije/dyw323

20. Keil AP, Edwards JK, Richardson DR, Naimi AI, Cole SR. The parametric G-formula for time-to-event data: towards intuition with a worked example. *Epidemiology*. 2014;25(6):889-897. doi:10.1097/EDE.0000000000000160

21. Robins JM, Greenland S, Hu FC. Estimation of the Causal Effect of a Time-Varying Exposure on the Marginal Mean of a Repeated Binary Outcome. *Journal of the American Statistical Association*. 1999;94(447):687-700. doi:10.1080/01621459.1999.10474168

22. Westreich D, Cole SR, Young JG, et al. The Parametric G-Formula to Estimate the Effect of Highly Active Antiretroviral Therapy on Incident AIDS or Death. *Stat Med*. 2012;31(18):2000-2009. doi:10.1002/sim.5316

23. Zivich PN, Ross RK, Shook-Sa BE, Cole SR, Edwards JK. Empirical Sandwich Variance Estimator for Iterated Conditional Expectation g-Computation. *Statistics in Medicine*. Published online November 3, 2024:sim.10255. doi:10.1002/sim.10255

24. Edwards JK, Cole SR, Zivich PN, Hudgens MG, Breger TL, Shook-Sa BE. Semiparametric g-computation for survival outcomes with time-fixed exposures: An illustration. *Annals of Epidemiology*. 2024;96:24-31. doi:10.1016/j.annepidem.2024.05.013

25. Eisenberg-Guyot J, Renson A. Competing classes confront competing risks: unraveling mortality inequities with parametric g-computation. *Am J Epidemiol*. 2025;194(8):2440-2444. doi:10.1093/aje/kwae417

26. Edwards JK, McGrath LJ, Buckley JP, Schubauer-Berigan MK, Cole SR, Richardson DB. Occupational Radon Exposure and Lung Cancer Mortality: Estimating Intervention Effects Using the Parametric g-Formula. *Epidemiology*. 2014;25(6):829-834. doi:10.1097/EDE.0000000000000164

27. Cole SR, Richardson DB, Chu H, Naimi AI. Analysis of Occupational Asbestos Exposure and Lung Cancer Mortality Using the G Formula. *American Journal of Epidemiology*. 2013;177(9):989-996. doi:10.1093/aje/kws343

28. Cole SR, Frangakis CE. The consistency statement in causal inference: a definition or an assumption? *Epidemiology*. 2009;20(1):3-5. doi:10.1097/EDE.0b013e31818ef366

29. Wen L, Young JG, Robins JM, Hernán MA. Parametric g-formula implementations for causal survival analyses. *Biometrics*. 2021;77(2):740-753. doi:10.1111/biom.13321

30. Robins J, Hernán M, Fitzmaurice G, Davidian M, Verbeke G, Molenberghs G. Longitudinal data analysis. *Handbooks of modern statistical methods*. Published online 2009:553-599.

31. Schomaker M, Luque-Fernandez MA, Leroy V, Davies MA. Using longitudinal targeted maximum likelihood estimation in complex settings with dynamic interventions. *Statistics in Medicine*. 2019;38(24):4888-4911. doi:10.1002/sim.8340

32. Rudolph JE, Schisterman EF, Naimi AI. A Simulation Study Comparing the Performance of Time-Varying Inverse Probability Weighting and G-Computation in Survival Analysis. *Am J Epidemiol*. 2023;192(1):102-110. doi:10.1093/aje/kwac162

33. Morris TP, White IR, Crowther MJ. Using simulation studies to evaluate statistical methods. *Statistics in Medicine*. 2019;38(11):2074-2102. doi:10.1002/sim.8086

34. Harris KM, Halpern CT, Whitsel EA, et al. Cohort Profile: The National Longitudinal Study of Adolescent to Adult Health (Add Health). *International Journal of Epidemiology*. 2019;48(5):1415-1415k. doi:10.1093/ije/dyz115

35. Gogtay NJ, Ranganathan P, Aggarwal R. Understanding estimands. *Perspectives in Clinical Research*. 2021;12(2):106. doi:10.4103/picr.picr_384_20

36. Perkins NJ, Cole SR, Harel O, et al. Principled Approaches to Missing Data in Epidemiologic Studies. *Am J Epidemiol*. 2018;187(3):568-575. doi:10.1093/aje/kwx348

37. Wen L, Marcus JL, Young JG. Intervention treatment distributions that depend on the observed treatment process and model double robustness in causal survival analysis. *Stat Methods Med Res*. 2023;32(3):509-523. doi:10.1177/09622802221146311

38. Zivich PN, Cole SR, Greifer N, Montoya LM, Kosorok MR, Edwards JK. Confidence Regions for Multiple Outcomes, Effect Modifiers, and Other Multiple Comparisons. *arXiv*. Preprint posted online 2025. doi:10.48550/ARXIV.2510.07076

**eAppendix**

**A. Identification for Semi-Competing Event Outcome with Time-Varying Actions**

Let $L_{i,k}, A_{i,k}, C_{i,k}, Y_{i,k}, Z_{i,k}$ be defined as in the main paper for unit $i$ and time interval $k \in \{1, \dots, \tau\}$. Also let $Y_{i,k}^{\bar{a}_{k-1}}$ and $Z_{i,k}^{\bar{a}_{k-1}}$ be similarly defined. Hereafter, we assume observations are independent and identically distributed so the index for units is left implicit. As before, interest is in the parameter vector,

$$\psi(\tau) = \begin{bmatrix} \psi_2(\tau) \\ \psi_3(\tau) \end{bmatrix} = \begin{bmatrix} E\left[I\left(Z_\tau^{\bar{a}_{\tau-1}} = 2\right)\right] - E\left[I\left(Z_\tau^{\bar{a}'_{\tau-1}} = 2\right)\right] \\ E\left[I\left(Z_\tau^{\bar{a}_{\tau-1}} = 3\right)\right] - E\left[I\left(Z_\tau^{\bar{a}'_{\tau-1}} = 3\right)\right] \end{bmatrix}$$

Now we can show that each component of $\psi(\tau)$ can be expressed in terms of the observed data. For ease of presentation, we first show identification without censoring (i.e., the observed data vector is $O_i = (L_0, A_0, Y_1, \dots, Y_{k-1}, L_k, A_k, Y_k)$). Further, identification is shown for $\tau \coloneqq 2$. For $z \in \{1,2,3\}$, it follows that

$$\begin{aligned} E\left[I\left(Z_2^{\bar{a}_1} = z\right)\right] &= E\left[E\left\{I\left(Z_2^{\bar{a}_1} = z\right) \mid L_0\right\}\right] \\ &= E\left[E\left\{I\left(Z_2^{\bar{a}_1} = z\right) \mid A_0 = a_0, L_0\right\}\right] \\ &= E\left[E\left\{E\left[I\left(Z_2^{\bar{a}_1} = z\right) \mid A_0 = a_0, \bar{L}_1\right] \mid A_0 = a_0, L_0\right\}\right] \\ &= E\left[E\left\{E\left[I\left(Z_2^{\bar{a}_1} = z\right) \mid A_0 = a_0, \bar{L}_1, Y_1^{a_0} \neq 3\right] \mid A_0 = a_0, L_0\right\}\right] \\ &= E\left[E\left\{E\left[I\left(Z_2^{\bar{a}_1} = z\right) \mid \bar{A}_1 = \bar{a}_1, \bar{L}_1, Y_1^{a_0} \neq 3\right] \mid A_0 = a_0, L_0\right\}\right] \\ &= E[E\{E[I(Z_2 = z) \mid \bar{A}_1 = \bar{a}_1, \bar{L}_1, Y_1 \neq 3] \mid A_0 = a_0, L_0\}] \end{aligned}$$

following from

1. The law of iterated expectation over $L_0$
2. Conditional exchangeability with positivity, $Z_2^{\bar{a}_1} \coprod A_0 \mid L_0$
3. The law of iterated expectation over $L_1$
4. We can separate $I\left(Z_2^{\bar{a}_1} = z\right)$ to $I\left(Z_1^{a_0} \neq 2\right) \times I\left(Z_2^{\bar{a}_1} = z\right)$. Since this equality will only be non-zero when $Z_1^{a_0} \neq 3$ we can add $Y_1^{a_0} \neq 3$ to the conditioning bar. Intuitively, we are restricting to those at "risk" of being in one of the 3 states at time 2.
5. Conditional exchangeability with positivity: $Z_2^{\bar{a}_1} \coprod A_1 \mid \bar{L}_1, A_0, Y_1^a \neq 3$
6. Causal consistency allows us to swap the potential outcomes with the observed outcomes

We now describe how this proof is extended for censoring and $\tau > 2$. First, censoring at $C_1 = 1$ can be freely conditioned on after Step 1 via an additional conditional exchangeability, $Z_2^{\bar{a}_1} \coprod C_1 \mid A_0 = a_0, L_0$, with positivity assumption. Similarly, $C_2 = 1$ can be conditioned on after Step 5 given the corresponding exchangeability and positivity assumptions provided in Table 1.

To extend this proof, for $\tau > 2$, note that analogs of Steps 3-5 can be iteratively applied to each inner expectation up to the desired time interval.

## B. Data generating mechanism

$L_{i,k}$ is a dichotomous, time-varying confounder that depends on prior $L$ and $A$. At baseline, $L_0$ is generated via

$$L_{i,0} \sim \text{Bernoulli}(0.5)$$

At $L_1$ and $L_2$, potential outcomes for $L$ are generated via

$$L_{i,1}(\bar{a}_0) \sim \text{Bernoulli}\left(\text{expit}\left(-1 - a_0 + L_{i,0}\right)\right)$$

$$L_{i,2}(\bar{a}_1) \sim \text{Bernoulli}\left(\text{expit}\left(-1 - a_1 + L_{i,1}(\bar{a}_0)\right)\right)$$

The outcome $Y_{i,k}$ follows the multistate structure previous established, where $Y_{i,k} = 1$ indicates no intermediate or terminal event at time $k$, $Y_{i,k} = 2$ indicates intermediate event, and $Y_{i,k} = 3$ indicates terminal event. Potential outcomes for $Y$ are generated from a multinomial distribution: $\frac{\exp(\eta_{i,j})}{\Sigma_{k=0}^{2}\exp(\eta_{i,k})}$ where $j \in \{1,2,3\}$. Linear function $\eta_{i,0}$= 1 (referent) and $\eta_{i,2}$ and $\eta_{i,3}$ are defined for each time point, $k$, via

$$k = 1:$$

$$\eta_{i,2} = 0.05 - 0.6a_0 - 2L_{i,0}$$

$$\eta_{i,3} = -1.75 - 0.6a_0 - 2L_{i,0}$$

$$k = 2:$$

$$\eta_{i,2} = 0.1 - 0.8a_1 - 2.2L_{i,1}(a_0) + 0.5Y_{i,1}(a_0)$$

$$\eta_{i,3} = -2 - 0.6a_1 - 2L_{i,1}(a_0) + 0.4Y_{i,1}(a_0)$$

$$k = 3:$$

$$\eta_{i,2} = 0.3 - 0.9a_2 - 2.2L_{i,2}(\bar{a}_1) + 0.5Y_{i,2}(\bar{a}_0)$$

$$\eta_{i,3} = -4 - 0.6a_1 - 2L_{i,2}(\bar{a}_1) + 0.4Y_{i,2}(\bar{a}_0)$$

Observed confounder and outcome data were generated from the potential outcomes with causal consistency, such that for a random variable, $X$,

$$X_{i,k+1} = \sum_{\bar{a}_k} X_{i,k+1}\,(\bar{a}_k) I(A_{i,k} = \bar{a}_k)$$

and observed exposure values are generated via

$$A_{i,0} \sim \text{Bernoulli}\left(\text{expit}\left(1 - 2L_{i,0}\right)\right)$$

$$A_{i,1} \sim \text{Bernoulli}\left(\text{expit}\left(-1 - L_{i,1} + 1.75A_{i,0}\right)\right)$$

$$A_{i,2} \sim \text{Bernoulli}\left(\text{expit}\left(-1 - L_{i,2} + 1.75A_{i,1}\right)\right)$$

Monotonic loss to follow-up is induced among those alive at each time via

$$C_{i,1} \sim \text{Bernoulli}\left(\text{expit}\left(-3 - 0.5A_{i,0}\right)\right)$$

$$C_{i,2} \sim \begin{cases} 1, & \text{if } C_{i,1} = 1 \\ \text{Bernoulli}\left(\text{expit}\left(-3 - 0.5A_{i,1}\right)\right), & \text{if } C_{i,1} = 0 \end{cases}$$

$$C_{i,3} \sim \begin{cases} 1, & \text{if } C_{i,2} = 1 \\ \text{Bernoulli}\left(\text{expit}\left(-3 - 0.5A_{i,2}\right)\right), & \text{if } C_{i,2} = 0 \end{cases}$$

where $L_{i,j}, A_{i,j}, Y_{i,j}$ are unobserved at all following time points if censored.

## C. Add Health details for illustrative example

eTable 1. Variable definitions for confounders included in the illustrative example.

| Variable | Variable type | Coding | Notes | Waves of measurement |
|---|---|---|---|---|
| Previous diagnosis of hypertension | Time-varying confounder | Dichotomous variable<br><br>1 = Self-reported previous diagnosis of hypertension;<br>0 = No previous diagnosis | | III, IV, V |
| Sex | Fixed confounder | Dichotomous variable<br><br>1 = male;<br>0 = female | If sex was missing at Wave I, substituted with non-missing data from the next available follow-up. | I |
| Race | Fixed confounder | Disjoint indicator terms<br><br>White, Black, Asian or, Pacific Islander, Other | White was the omitted term. If race was missing at Wave I, substituted with non-missing data from the next available follow-up. | I |
| Hispanic Ethnicity | Fixed confounder | Dichotomous variable<br><br>1 = Hispanic;<br>0 = Non-Hispanic | If ethnicity was missing at Wave I, substituted with non-missing data from the next available follow-up. | I |
| Age at W1 | Fixed confounder | Disjoint indicator terms | Age 16 was the omitted term. | I |
| | | 13, 14, 15, 16, 17, 18 | | |
| Weekly alcohol consumption | Time-varying confounder | Dichotomous variable<br><br>1 = Weekly or more frequent consumption of at least one alcoholic;<br>0 = Less than weekly consumption | | III, IV, V |
| Self-rated health | Time-varying confounder | Disjoint indicator terms<br><br>Excellent, very good, good, fair/poor | Fair and poor categories were collapsed due to sparsity. Excellent health was the omitted term. | III, IV, V |
| Exercise frequency | Time-varying confounder | Dichotomous variable<br><br>1 = 7+ instances of exercise in last 7 days;<br>0 = less than 7 instances. | Number of instances of exercise in last 7 days assessed in activity-specific questions.[a] Instances were summed across activities. Coded into a dichotomous variable due to variable distribution. | III, IV, V |
| Health insurance coverage | Time-varying confounder | Dichotomous variable<br><br>1 = currently has health insurance;<br>0 = does not have health insurance. | Wave III health insurance question asked the number of months participants had health insurance in last 12 months. We considered a participant to have health insurance if they reported 12 of 12 months insured. Waves IV-V asked how participants received health insurance. We considered a participant to have health insurance they reported any type of insurance (from employer, bought on marketplace, veteran's benefits, etc.). | III, IV, V |
| Educational attainment | Time-varying confounder | Dichotomous variable<br><br>1 = Completed college education or higher education;<br>0 = Some college or less. | Educational attainment was not assessed at Wave III because most participants were not expected to have completed college yet (ages 18-26). | IV, V |
| Height | Time-varying confounder | Continuous | Splines were determined separately for each time-point | III, IV, V |
| | | Restricted quadratic splines | based on visualization of loess plot.<br><br>For each wave, if study measured height was refused or missing, self-reported height was substituted if available. Height was converted to centimeters for consistency across measurements.<br>Splines were determined separately for each time-point based on visualization of loess plot. | |
| Weight | Time-varying confounder | Continuous<br><br>Restricted quadratic splines | For each wave, if study measured weight was refused or missing, self-reported height was substituted if available. If measured weight exceeded study scale limits of 330 lbs or 149 kg, self-reported weight was substituted if available and over 330 lbs/149 kg, otherwise, weight was coded as 149 kg. Weight was converted to kilograms for consistency across measurements. | III, IV, V |
| Ever tried a cigarette | Fixed confounder | Dichotomous variable<br><br>1 = Ever tried a cigarette in the past;<br>0 = Never tried a cigarette. | | III |
| Underlying heart condition | Exclusion criteria | Dichotomous variable<br><br>1 = Heart condition;<br>0 = No heart condition. | Indicated if participant listed taking prescription medication for a heart problem in the past 12 months. | III |

[a] Activity-specific exercise questions: "In the past 7 days, how many times did you…"

- "Bicycle, skateboard, dance, hike, hunt, or do yard work?"
- "Roller blade, roller skate, downhill ski, snow board, play racquet sports, or do aerobics?"
- "Participate in gymnastics, weightlifting, or strength training?"
- "Participate in an individual sport such as running, wrestling, swimming, cross-country skiing, cycle racing, martial arts?"

- "Participate in a in a strenuous team sport such as football, soccer, basketball, lacrosse, rugby, field hockey, or ice hockey?" [Wave V combines individual and team sport questions]
- "Play golf, go fishing or bowling, play softball or baseball?"
- "Walk for exercise?"

All confounders were included in models for baseline and two follow-up times, $k \in \{0, 1, 2\}$, except for education attainment which was included at $k$ = 1 and 2 and ever tied a cigarette which was only included at $k$ = 0. No interaction terms were included in the models.

eTable 2. Characteristics of study sample by follow-up wave; Add Health Cohort, 1996-2024.

| **Characteristic** | **Wave III** (ages 18-26) *n* = 13,909 | **Wave IV** (ages 24-32) *n* = 11,563 | **Wave V** (ages 33-43) *n* = 3,951 | **Wave VI** (ages 39-51) *n* = 2,954 |
|---|---|---|---|---|
| Age at Wave I, mean (SD)* | 15.61 (1.58) | 15.59 (1.58) | 15.51 (1.59) | 15.5 (1.59) |
| Male, *n* (%)* | 6563 (47.2%) | 5292 (45.8%) | 1581 (40%) | 989 (38.1%) |
| Race, *n* (%)* | | | | |
| White | 7741 (55.7%) | 6675 (57.7%) | 2588 (65.5%) | 1742 (67.2%) |
| Black | 2968 (21.3%) | 2429 (21%) | 734 (18.6%) | 444 (17.1%) |
| Asian | 1010 (7.3%) | 772 (6.7%) | 215 (5.4%) | 143 (5.5%) |
| Other | 2190 (15.7%) | 1687 (14.6%) | 414 (10.5%) | 414 (10.5%) |
| Hispanic, *n* (%)* | 2202 (15.8%) | 1716 (14.8%) | 456 (11.5%) | 296 (11.4%) |
| College attainment | - | 1522 (13.3%) | 1898 (49.7%) | 1408 (54.7%) |
| Ever tried smoking a cigarette, *n* (%) | 8177 (58.8%) | - | - | - |
| Current cigarette smoker, *n* (%) | - | 4001 (34.8%) | 844 (22.1%) | 389 (15.1%) |
| Weekly alcohol consumption, *n* (%) | 3882 (27.9%) | 3463 (30.1%) | 669 (37.2%) | 1333 (51.8%) |
| Exercise 7+ times over previous 7 days, *n* (%) | 5144 (37%) | 4578 (39.9%) | 1698 (44.4%) | 1745 (67.8%) |
| Self-rated health, *n* (%) | | | | |
| Excellent | 4624 (33.2%) | 2210 (19.2%) | 690 (18.1%) | 317 (12.3%) |
| Very Good | 5629 (40.5%) | 4424 (38.5%) | 1489 (39%) | 903 (35.1%) |
| Good | 3034 (21.8%) | 3770 (32.8%) | 1204 (31.5%) | 969 (37.7%) |
| Fair/Poor | 622 (4.5%) | 1082 (9.4%) | 438 (11.5%) | 384 (14.9%) |
| Health insurance, *n* (%) | 8844 (63.6%) | 9213 (80.2%) | 3575 (93.6%) | 2447 (95.1%) |
| Height in cm, mean (SD) | 169.91 (10.46) | 169.82 (10.02) | 170.04 (9.71) | 169.91 (9.57) |
| Weight in kg, mean (SD) | 76.85 (20.71) | 83.93 (23.04) | 87.21 (22.61) | 89.65 (22.97) |
| Previous diagnosis of hypertension, *n* (%) | 768 (5.5) | 1175 (10.2) | 679 (17.8) | 596 (23.2) |
| Hypertensive, *n* (%) | - | 2090 (18.1%) | 1421 (37.2%) | 459 (17.7%) |
| Death, *n* (%) | - | 77 (0.7%) | 130 (3.3%) | 21 (0.8%) |

eTable 3. Results at end of follow-up for applied example comparing effect of cigarette smoking prevention to the naturally observed smoking pattern from early adulthood to mid-adulthood on prevalent hypertension and incident death in mid-adulthood using Add Health.

| **Estimator** | **All cigarette smoking prevented % (95% CI)** | **Naturally occurring smoking pattern % (95% CI)** | **Effect of intervention % (95% CI)[a]** |
|---|---|---|---|
| Alternative estimator 1[b] | | | |
| Hypertension | 16.0 (14.4, 17.6) | 16.4 (15.0, 17.9) | -0.4 (-1.4, 0.5) |
| Death | 5.8 (4.8, 6.7) | 8.1 (7.2, 9.1) | -2.4 (-3.1, -1.6) |
| Alternative estimator 2[c] | | | |
| Hypertension | 19.4 (17.6, 21.3) | 21.0 (19.2, 22.8) | -1.6 (-2.7, -0.5) |
| Death | 0 (0, 0) | 0 (0, 0) | 0 (0, 0) |
| ICE g-computation | | | |
| Hypertension | 18.4 (16.6, 20.1) | 19.5 (17.8, 21.1) | -1.1 (-2.1, -0.1) |
| Death | 3.9 (3.1, 4.7) | 5.5 (4.7, 6.3) | -1.6 (-2.3, -1.0) |

[a] Prevalence difference for hypertension and risk difference for death.

[b] G-computation estimator that accounts for semi-competing risks but only estimates the effect of the action at baseline.

[c] ICE g-computation estimator that accounts for time-varying confounding but treats terminal event as censoring (no estimation of terminal event risk).